\documentclass[12pt]{JHEP3}

\usepackage[pdftex]{graphicx}
\newcommand\fverb{\setbox\pippobox=\hbox\bgroup\verb}
\newcommand\fverbdo{\egroup\medskip\noindent%
\fbox{\unhbox\pippobox}\ }			
\newcommand\fverbit{\egroup\item[\fbox{\unhbox\pippobox}]}
\newbox\pippobox
\def\d2bar{$\overline{\mbox D2}$}
\title{The Entropy Function for \\the Black Holes of Nariai Class}
\author{Jin-Ho Cho$^{1,2}$ and Soonkeon Nam$^{1}$\\
$^{1}$Department of Physics \& Research Institute for Basic Sciences,\\ Kyung Hee University, Seoul 130-701, Korea\\
$^{2}$Center for Quantum Space Time, Sogang University, Seoul 121-742, Korea\\\\
E-mail: \email{cho.jinho@gmail.com}, \email{nam@khu.ac.kr}}

\preprint{arXiv{: 0711.2514}} 

\abstract{Based on the fact that the near horizon geometry of the extremal Schwarzschild-de Sitter black holes is Nariai geometry, we define the black holes of Nariai class as the configuration whose near-horizon geometry is factorized as two dimensional de Sitter space-time and some compact topology, that is Nariai geometry. We extend the entropy function formalism to the case of the black holes of Nariai class. The conventional entropy function (for the extremal black holes) is defined as Legendre transformation of Lagrangian density, thus the `Routhian density', over two dimensional anti-de Sitter. As for the black holes of Nariai class, it is defined as {\em minus} `Routhian density' over two dimensional de Sitter space-time. We found an exact agreement of the result with Bekenstein-Hawking entropy. The higher order corrections are nontrivial only when the space-time dimension is over four, that is, $d>4$. There is a subtlety as regards the temperature of the black holes of Nariai class. We show that in order to be consistent with the near horizon geometry, the temperature should be non-vanishing despite the extremality of the black holes. }  


\begin{document}

\section{Introduction}
The entropy function formalism of Sen is a neat way to compute the entropy of an extremal black hole even without the details of the solution \cite{Sen:2005wa,Sen:2005iz}. This method is especially very useful when we are interested in the entropy contribution coming from the higher order corrections to the Einstein-Hilbert action, as can be expected from string theory. The only necessary information about an extremal black hole is its near horizon geometry that takes the form, AdS$_{2}\times K^{d-2}$, where AdS$_2$ stands for two dimensional anti-de Sitter space-time while $K^{d-2}$ is some $(d-2)$-dimensional compact manifold. In this regard, the formulation incorporates the attractor property of the black hole from the beginning \cite{Alishahiha:2006jd,Dabholkar:2006tb,Cardoso:2006xz,Garousi:2007zb,Cai:2007an,Cho:2007mn,Astefanesei:2007bf,Myung:2007an}. (See also Ref. \cite{Sen:2007qy} and references therein.) The fact that the formalism  successfully reproduces Bekenstein-Hawking entropy suggests that the entropy is not sensitive to the asymptotic behaviors of various fields involved.  

This attractor behavior is mainly due to the long throat structure \cite{Kallosh:2006bt}. The fields run through an infinite throat region to reach their attractor values and forget their initial ones. This suggests that the entropy function formalism could also be applicable to the cases with the near-horizon geometries other than AdS$_{2}\times K^{d-2}$: So far as there is a Freund-Rubin type compactification \cite{Freund:1980xh}, thereby making an infinitely long throat region near the horizon, then the fields could show similar attractor behavior.  On the other hand, the near-horizon isometry does an important role in regulating the forms of most fields. The more symmetries we have, the more concretely the forms of the fields will be determined. 

As a simple extension of the entropy function formalism, one could conceive the cases which entail, near their horizons, two dimensional de Sitter space-time (dS$_2$), that is, another familiar symmetric space. In this regard, we have two instant questions. First, is there at all any black hole that possesses de Sitter space-time as the near-horizon geometry? In principle, this looks possible as long as the trace $T\equiv T^{\mu}{}_{\mu}$ of the energy-momentum tensor of the matter fields and the cosmological constant $\Lambda_{d}$ are appropriately chosen. By taking trace over the Einstein equation one gets
\begin{equation} 
R=4\Lambda_{d}-16\pi G_{d}T,
\end{equation}   
where $G_{d}$ is $d$-dimensional Newton constant.
For the geometry factorized into dS$_{2}\times K^{d-2}$,
the curvature scalar $R$ is positive. Therefore one necessary condition for the specific factorization of the the geometry would be $\Lambda_{d}>4\pi G_{d}T$. 

The second question is a bit technical one. What would be the expression for the entropy function of the black holes which contain dS$_{2}$ near the horizons? 
The entropy function of the conventional extremal black holes can be understood as the Routhian density over two-dimensional anti-de Sitter space-time. If we just extrapolate the definition to the cases we are interested in now, it might result in some negative entropy. For the geometry without the angular momentum, the Einstein-Hilbert term will contribute to the entropy $S$ in the form
\begin{eqnarray} 
S&\sim&-\frac{1}{16\pi G_{d}}\int_{K} d\Omega\, \left(R-2\Lambda_{d} \right)+ \cdots  \nonumber\\
&=&-\frac{1}{16\pi G_{d}}\int_{K} d\Omega\, \left(2\Lambda_{d}-16\pi G_{d}T \right)+\cdots.
\end{eqnarray}      
This is the value on shell and the dots stand for the contribution from other matter fields.
Hence it could be negative if $T\le0$ and the cosmological constant is sufficiently large to dominate over other contributions. 

The aim of this paper is to answer the above two questions.
We will show an explicit example of the black holes with de Sitter near-horizon geometry. In the example, $\Lambda_{d}>0$ and $T=0$, therefore the necessary condition of $\Lambda_{d}> 4\pi G_{d}T$ is satisfied. Actually such type of factorization of the near-horizon geometry into the form dS$_{2}\times K^{d-2}$ is generic  whenever the metric function is negative near its double zero.  

We will also see that the entropy function for the black holes with de Sitter near-horizon geometry is defined as the minus Routhian density over the de Sitter part, that is, $S=-2\pi H$. Despite the negative value of the Routhian $H$, the entropy is thus positive. 

This paper is organized as follows. In the next section, we consider an extremal Schwarzschild-de Sitter black hole in $4$-dimensions. We discuss its global structure and some of its thermodynamic properties. In Sec. \ref{seciii}, we show that the near-horizon geometry of the extremal Schwarzschild-de Sitter black hole is factorized into dS$_{2}\times$S$^{2}$, that is into Nariai geometry. Based on this observation, we define the black holes of Nariai class in general $d$-dimensions as the black holes whose near-horizon geometries contain two dimensional de Sitter space-time. In Sec. \ref{seciv}, we derive the entropy function, {\it \`{a} la} Sen \cite{Sen:2005wa}, starting from Wald's entropy formula \cite{Wald:1993nt,Jacobson:1993vj,Iyer:1994ys,Jacobson:1994qe}.  Sec. \ref{secv} discusses the entropy contribution coming from the higher derivative corrections to the Einstein-Hilbert action. In $4$-dimensions, Gauss-Bonnet term contributes a constant addition to the entropy. Sec. \ref{secvi} discusses the difference of Nariai geometry discussed in this paper from the ones appearing in the region between the event horizon and the cosmological horizon of Schwarzschild-de Sitter black holes in the extremal limit. We also discuss the issue of the temperature raised specifically in the black holes of Nariai class. 
We argue that the temperature of the black holes of Nariai class is not zero despite their extremality.

\section{Basics of Extremal Schwarzschild-de Sitter Black Holes}\label{secii}
In de Sitter background, Schwarzschild black holes can be extremal possessing degenerate horizon.
The geometry of Schwarzschild-de Sitter black holes is given as follows:
\begin{eqnarray}\label{q8}
ds^{2}&=&-f(r)dt^{2}+ \frac{1}{f(r)}dr^{2}+ r^{2}d\Omega^{2}_{2}, \nonumber\\
f(r)&=&1- \frac{2G_{4}M}{r}-\frac{r^{2}}{l^{2}},
\end{eqnarray}
where $G_{4}$ is $4$-dimensional Newton's constant and $l$ is the length scale characterizing the cosmological constant, that is,
\begin{equation}\label{q9}
\frac{1}{l^{2}}= \frac{\Lambda_{4}}{3}.
\end{equation}


\FIGURE{
\centering
\includegraphics[width=9 cm]{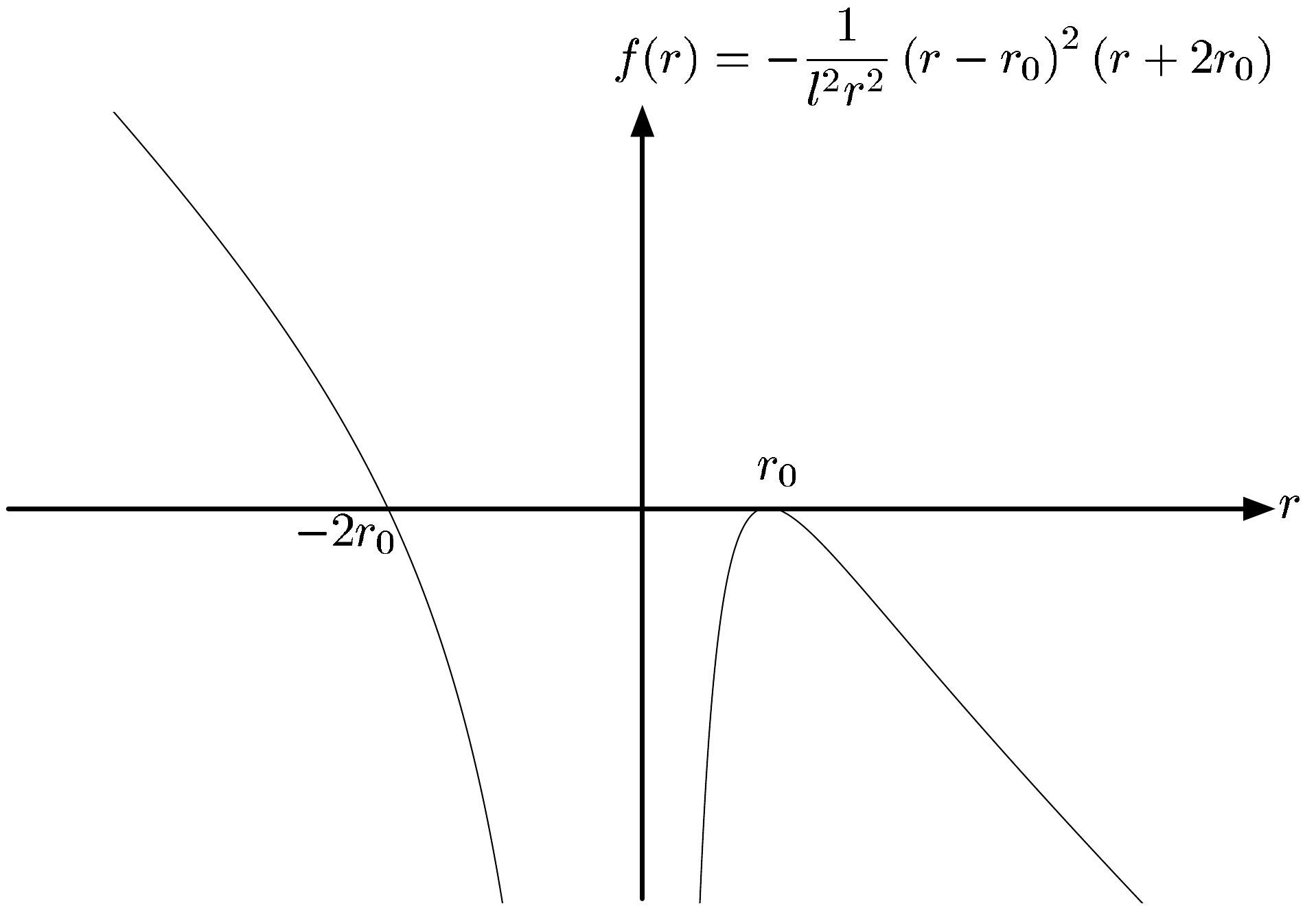}
\caption{\small The extremal Schwarzschild-de Sitter metric has one degenerate horizon (at $r=r_0$) and another unphysical horizon (at $r=-2r_0$). The metric function $g^{rr}=f(r)$ is negative around its double zero.}
\label{fig0}
}

The zeros of the metric component, $g^{rr}=f(r)$, signify the event horizons of a black hole. The surface of  constant $r$ is null at the zeros of the metric function $f(r)$.
The presence of the term concerning the cosmological constant in the metric function $f(r)$, allows the  extremal case, despite the neutrality of the black hole. We can factorize the function $f(r)$ as follows:
\begin{equation}\label{factor}
f(r)=- \frac{1}{l^{2}r} \left(r-r_0 \right)^{2} \left( r+ 2r_0 \right). 
\end{equation} 
See Fig. \ref{fig0} for the form of the metric function. 
In this specification, the double zero is related with the parameters $M$ and $l$ as
\begin{eqnarray}
&&r_0^{3}=G_{4}Ml^{2},\label{20b} \\
&&r_0^{2}=\frac{l^{2}}{3}\label{20c}, 
\end{eqnarray}
which implies $r_0=3G_{4}M$ and the BPS like equation relating the mass parameter and the cosmological constant; 
\begin{equation} 
27G_{4}^{2}M^{2}=l^{2}.
\end{equation}  

One thing to note is that the metric function $f(r)$ is negative at every point $r$($>0$) other than $r_0$. This means that the coordinate $r$ ($0\leq r<\infty$) is the temporal coordinate in most region except the point $r_0$, where it becomes the null coordinate. In the meantime, the coordinate $t$ ($-\infty<t<\infty$) is now one of the spatial coordinates. The geometry, being dependent only on the temporal coordinate $r$, is neither static nor stationary. 

Various properties concerning the global structure of the extremal Schwarzschild-de Sitter black hole were worked out in Ref. \cite{Podolsky:1999ts}. Its result can be summarized as the Penrose diagram shown in Fig. \ref{fig1}. Generic observer going through the horizon at the instant $r=r_0=3G_{4}M$ is destined to the singularity at the future infinity $r=0$. 


\FIGURE{
\centering
\includegraphics[width=11 cm]{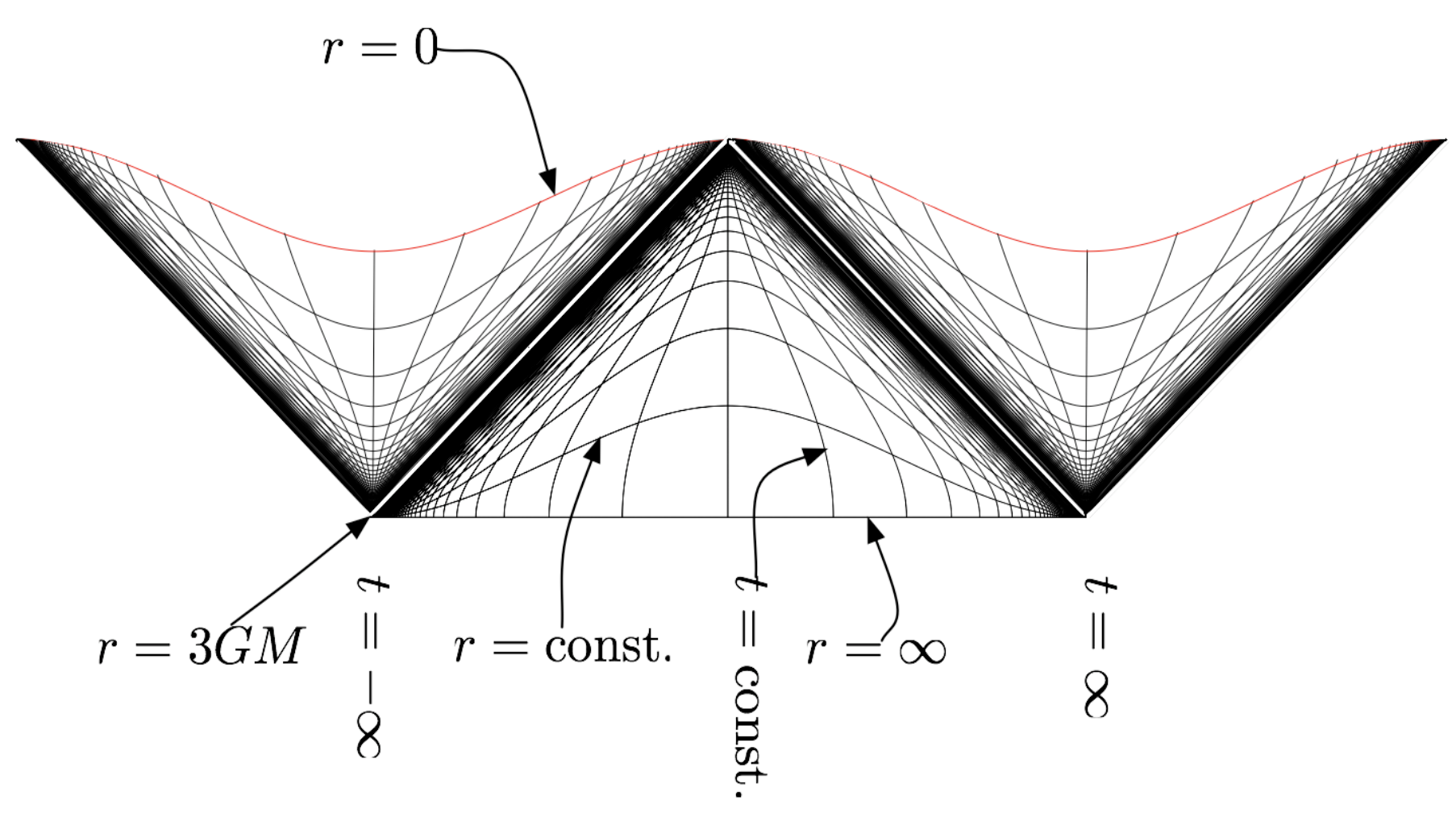}
\caption{\small Penrose diagram of an extremal Schwarzschild-de Sitter black hole. The geometry is singular at $r=0$.}
\label{fig1}
}

The surface gravity (given by $\kappa\equiv \vert f'(r_{0})/2\vert$) of this extremal black hole vanishes because the metric function has a double zero at the degenerate horizon. However, one should take this definition of the surface gravity with a grain of salt. The above definition of the surface gravity can be recast in terms of a Killing vector $t^{\mu}=dx^{\mu}/dt$ as
\begin{equation} 
\left.\kappa^{2}=-\frac{1}{2}t^{\mu;\nu}t_{\mu;\nu}\right\vert_{r=r_{0}},
\end{equation}
where $t_{\mu;\nu}$ stands for the covariant derivative of the vector $g_{\mu\rho}t^{\rho}$ with respect to the coordinate $x^{\nu}$. 
Though the Killing vector $t^{\mu}$ is normalized in the asymptotically flat region in the limit of $l \rightarrow \infty$,  there is no asymptotically flat region in the generic de Sitter background. One may argue that the surface gravity be defined with respect to an observer following `geodesic orbit' \footnote{the geodesic line on which the Killing vector $t^{\mu}$ is tangential} and who feels no acceleration \cite{Bousso:1996au}. In this scheme, one has to replace $t^{\mu}$ with 
\begin{equation} 
k^{\mu}=\frac{1}{\sqrt{-f(r_{g})}}t^{\mu}
\end{equation}  
that is normalized at $r=r_{g}$ on the geodesic orbit. The modified surface gravity at the horizon $r=r_{h}$ satisfies
\begin{equation}\label{modified}
\tilde{\kappa}^{2}=\pm \frac{\left.f'^{2}(r)\right\vert_{r=r_{h}}}{4 f(r_{g})},
\end{equation}  
where the upper sign applies to the region where $f(r)>0$ while the lower sign is for the case at hand. The problem with the extremal Schwarzchild-de Sitter black holes is that $r_{g}=r_{h}=r_{0}$ and therefore $f(r_{g})=0$ because the geodesic orbit is determined by $f'(r_{g})=0$. One way out for the case at hand would be to define it as
\begin{equation} 
\tilde{\kappa}^{2}=\pm\lim_{r \rightarrow r_{0}}\frac{f'^{2}(r)}{4 f(r)}=\pm \frac{f''(r_{0})}{2},
\end{equation}  
which results in $\tilde{\kappa}^{2}=3/l^{2}$ for the extremal Schwarzschild-de Sitter black hole.
Zero surface gravity is the result obtained by extrapolating the surface gravity defined in the asymptotically flat space-time.

Bekenstein-Hawking entropy of the black hole can be read from $r_0^{2}$ as
\begin{equation}\label{20.9}
S=\frac{4\pi r_0^{2}}{4G_{4}}= \frac{\pi l^{2}}{3G_{4}}=9\pi G_{4}M^{2}. 
\end{equation}  

\section{Black Holes of Nariai Class}\label{seciii}

Since Nariai first found a cosmological solution of the type dS$_{2}\times$S$^{2}$ (thus named as Nariai geometry) in four dimensional de Sitter background \cite{nariai}, Ginsparg and Perry realized that the same geometry appears between two horizons of Schwartzschild-de Sitter black hole in the extremal limit of merging those two horizons, that is, the black hole horizon and the cosmological horizon \cite{Ginsparg:1982rs}. Further elaboration and its extension to the charged Nariai geometry were made by Bousso and Hawking \cite{Bousso:1996au}.  See also Refs. \cite{Bousso:1996pn,Dias:2003up,Cardoso:2004uz,Diaz:2007ts} for other extensions of Nariai geometry. 

In this paper, we will consider a different case and see below that Nariai solution appears also as the geometry near the region trapped by two merged horizons. In other words, it appears as the near horizon geometry (the geometry near the time $r=r_0$) of the extremal Schwarzschild-de Sitter black holes. 

It is convenient to introduce the following near-horizon coordinates;
\begin{equation} 
\tau=\frac{r-r_0}{\epsilon},\qquad x=\epsilon\, t,\qquad (\epsilon\ll 1)
\end{equation}  
in terms of which the metric of (\ref{q8}) becomes factorized as dS$_{2}\times$S$^{2}$, that is, two-dimensional de Sitter space-time and a two-dimensional sphere;
\begin{equation} 
ds^{2}=- \frac{l^{2}}{3\tau^{2}}d\tau^{2}+ \frac{3\tau^{2}}{l^{2}}dx^{2}+r_0^{2}d\Omega^{2}_{2}.
\end{equation}  
The coordinate $\tau$ ranges from $-\infty$ to $\infty$ and its positive value corresponds to a point exterior to the degenerate horizon.
Both dS$_{2}$ and S$^{2}$ are of the same size $r_{0}=l/\sqrt{3}$, where $l$ is the size of the embedding $(3+1)$-dimensional de Sitter background. This specific geometry coincides with the one found by Nariai. In the near horizon region, the spatial section of the geometry has the topology of a hyper-cylinder, that is, $\mathbf{ R}\times$S$^{2}$. Especially one cannot see the black hole singularity in a finite time $\tau>-\infty$$ (r=0)$ and has only the accelerating horizon waiting for her at the time $\tau=0$ $(r=r_{0})$. 

One can relate the above near-horizon coordinates $(\tau,\, x)$ with the conventional planar coordinates by the relations;
\begin{eqnarray} 
\tau&=&\frac{l}{\sqrt{3}}e^{-\xi},\,\quad x=\frac{l}{\sqrt{3}}y \qquad (0<\tau<\infty)\nonumber\\
\tau&=&-\frac{l}{\sqrt{3}}e^{\lambda},\quad x=\frac{l}{\sqrt{3}}y \qquad (-\infty<\tau<0).
\end{eqnarray}
Then the geometry is described by
\begin{equation}\label{planar}
\frac{3}{l^{2}}ds^{2}=\left\{\begin{array}{cc}-d\xi^{2}+e^{-2\xi}dy^{2} & \quad(\tau>0) \\ &  \\ -d\lambda^{2}+ e^{2\lambda}dy^{2} & \quad(\tau<0).\end{array}\right.
\end{equation}  
Fig. \ref{desitterpenrose} shows the corresponding Penrose diagrams. The horizon is at the time $\xi=\infty$ and $\lambda=-\infty$.


\FIGURE{
\centering
\includegraphics[width=7 cm]{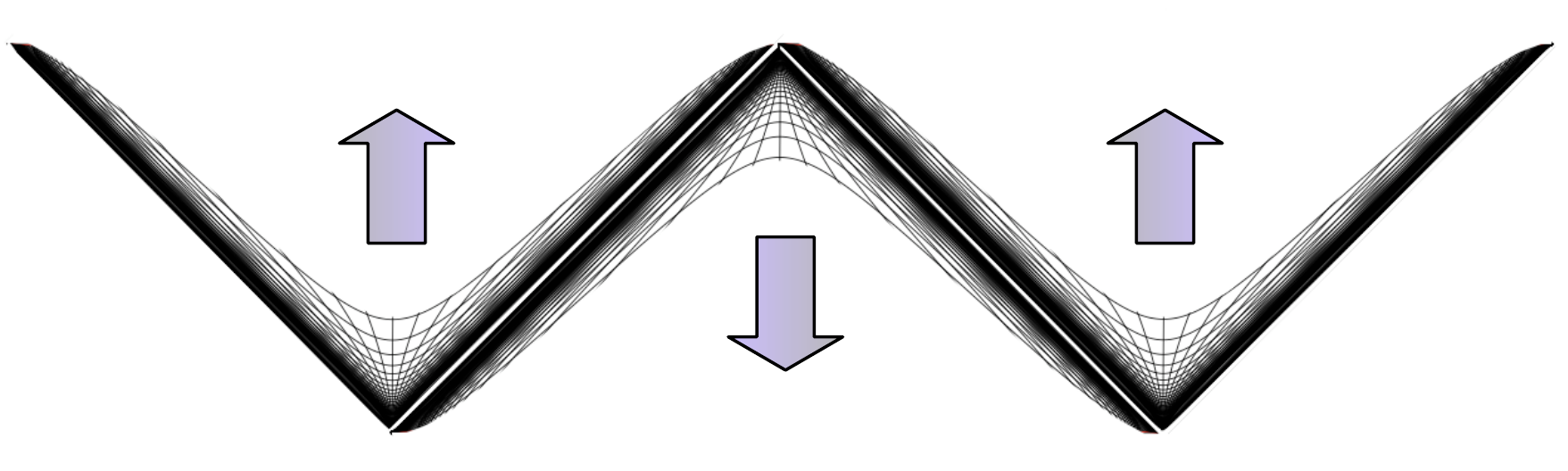}
\includegraphics[width=7 cm]{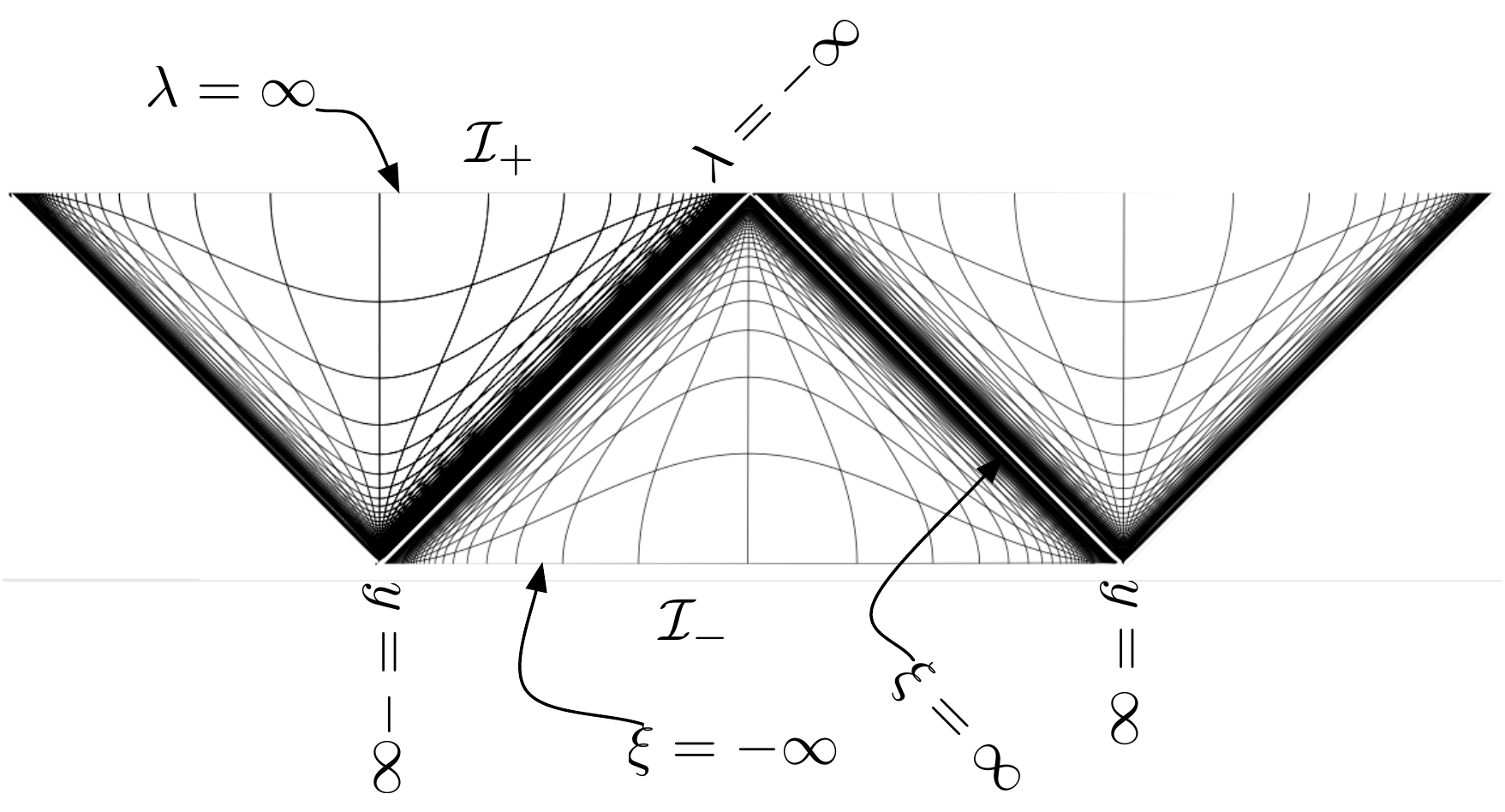}
\caption{\small The near-horizon region (left) can be expanded to describe $2$-dimensional de Sitter space-time (right). Every point in the right diagram corresponds to a two sphere of constant radius $r_{0}=l/\sqrt{3}$ even at the horizon $\xi=\infty$ and $\lambda=-\infty$. This feature is distinct from that of $4$-dimensional de Sitter.}
\label{desitterpenrose}
}

Based on this simple observation about the near-horizon Nariai geometry of the extremal Schwarzschild-de Sitter black holes, we set up the following ansatz for the black hole of Nariai class in $d$-dimensions. It is the black hole whose near-horizon geometry is factorized as $2$-dimensional de Sitter space-time and a $(d-2)$-dimensional sphere;
\begin{equation}\label{ansatz1}
ds^{2}=v_{1} \left(-\frac{d\tau^{2}}{\tau^{2}}+\tau^{2}dx^{2} \right)+v_{2}\,d\Omega^{2}_{d-2}. 
\end{equation}  
Being the symmetric spaces, de Sitter space-time and the sphere have the simple forms of the curvature tensor;
\begin{equation}\label{curvature}
R_{\alpha\beta\gamma\delta}= \frac{1}{v_{1}}\left(g_{\alpha\gamma}g_{\beta\delta}-g_{\alpha\delta}g_{\beta\gamma} \right),\qquad R_{\mu\nu \rho\sigma}=\frac{1}{v_{2}} \left(g_{\mu\rho}g_{\nu\sigma}-g_{\mu\sigma}g_{\nu\rho} \right).  
\end{equation}  

There might be other fields like scalars and various form fields. The only thing constraining these fields is that they respect the isometry group SO$(2,1)\times$SO$(d-1)$. In other words, the fields should be Lie invariant with respect to Killing vectors concerning the isometry;
\begin{eqnarray}\label{ansatz2}
\phi^{i}&=&u^{i} \nonumber\\
F^{j}_{e}&=&e^{j}\,d\tau\wedge dx,\qquad F^{i}_{m}=\frac{p^{j}}{\mbox{Vol}(\mbox{S}^{d-2})}\,d\Omega_{d-2}.
\end{eqnarray}

We have to note that the factor SO$(2,1)$ in the isometry group is $(1+1)$-dimensional de Sitter group and is not to be confused with $(1+1)$-dimensional anti-de Sitter group, that is, SO$(1,2)$. These two groups are defined as the groups which leave the hypersurface satisfying $x^{2}\pm y^{2}-z^{2}=\pm 1$ with the upper signature for de Sitter and the lower one for anti-de Sitter. There is no mathematical difference but there is physical distinction between them. $(1+1)$-dimensional de Sitter space-time is the hyper-surface embedded into $(2+1)$-dimensional space-time while $(1+1)$-dimensional anti-de Sitter space-time hyper-surface is embedded into $(1+2)$-dimensional space-time that has one spatial direction and two temporal directions.

\section{The Entropy Function}\label{seciv}

In this section, we develop the entropy function formalism for the black holes of Nariai class.
The entropy function as was defined in Ref. \cite{Sen:2005wa} is nothing but Wald's entropy formula \cite{Wald:1993nt,Jacobson:1993vj,Iyer:1994ys,Jacobson:1994qe},
\begin{equation}\label{waldformula}
S_{BH}=-8\pi\int_{\mbox{horizon}}d\theta d\varphi \frac{\delta{\mathcal S}}{\delta R_{x \tau x\tau}}\sqrt{-g_{\tau\tau}g_{xx}}
\end{equation}  
applied to the spherically symmetric extremal black holes. The result is that the entropy function is `Routhian density' over two dimensional anti-de Sitter space-time. Regarding the black holes of Nariai class, we have to modify the definition of the entropy function as {\em minus} `Routhian density' over two dimensional de Sitter space-time. Below, we just follow the procedure of Ref. \cite{Sen:2005wa}, that will lead us to this conclusion.

As for those field configurations compatible with the near-horizon isometry, the action ${\mathcal S}$ is just an algebraic polynomials of those field strengths (without any derivative involved). The functional derivative in Eq. (\ref{waldformula}) becomes simplified as the ordinary derivative;
\begin{equation} 
\frac{\delta{\mathcal S}}{\delta R_{x\tau x\tau}}=\frac{\partial {\mathcal L}}{\partial R_{x\tau x\tau}}\sqrt{-g}
\end{equation}  
resulting in
\begin{equation}\label{entropy}
S_{BH}=8\pi A \frac{\partial {\mathcal L}}{\partial R_{x\tau x\tau}}g_{\tau\tau}g_{xx}=-8\pi A \frac{\partial {\mathcal L}}{\partial R_{x\tau x\tau}}v^{2}_{1}.
\end{equation}  
Here, $A$ stands for the area of the horizon. 

On the other hand, the above entropy can be written in terms of Lagrangian density over two dimensional de Sitter space-time;
\begin{equation} 
L(\vec{e},\vec{p},\vec{u},\vec{v})=\int_{\mbox{S}^{d-1}}d\Omega_{d-1}\sqrt{-g}\,\mathcal{L}
\end{equation} 
Following the prescription of Ref. \cite{Sen:2005wa}, we multiply a parameter $\lambda$ on every occurrence of the curvature tensor $R_{\alpha\beta\gamma\delta}$, in other words, we replace the curvature tensor with $\lambda R_{\alpha\beta\gamma\delta}$, to define the function $L_{\lambda}(\vec{e},\vec{p},\vec{u},\vec{v})$. Then it is easy to see that
\begin{eqnarray} 
\left.\frac{\partial L_{\lambda}}{\partial \lambda}\right\vert_{\lambda=1}&=&\int_{\mbox{S}^{d-1}}d\Omega_{d-1}\sqrt{-g}\left( -2v^{2}_{1}R\right)\frac{\partial \mathcal{L}}{\partial R_{ x\tau x\tau}} \nonumber\\
&=&-4v^{2}_{1}A\frac{\partial \mathcal{L}}{\partial R_{ x\tau x\tau}}.
\end{eqnarray}  
The difference from the conventional extremal case lies in the expression for the curvature scalar $R$ used in the second line. As for Nariai case, we use $R=2/v_{1}$ of de Sitter rather than $R=-2/v_{1}$ of anti-de Sitter. Therefore we can represent the entropy of (\ref{entropy}) as
\begin{equation}\label{entropy2}
S_{BH}=2\pi \left.\frac{\partial L_{\lambda}}{\partial \lambda}\right\vert_{\lambda=1}.
\end{equation}  
Note that the case of the conventional extremal black holes comes with the opposite sign of the value on the right.

The remaining procedure of rewriting the right hand side as `Routhian density' can be followed after Ref. \cite{Sen:2005wa}. We just summarize here a few key steps developed there. The partial derivative of $L_{\lambda}$ with respect to $\lambda$ is related to other derivatives of the same function $L_{\lambda}$ with respect to the fields $\vec{u},\, \vec{v},\, \vec{e},\, \vec{p}$. We have invariance of the Lagrangian density under reparametrization of $x$ and $\tau$ coordinates.
As for the curvature, every factor of the Riemann tensor component, $R_{x\tau x\tau}$ should appear as the combination 
$\lambda g^{xx}g^{\tau\tau} R_{x\tau x\tau}= \lambda v_1^{-1}$. As for the gauge field, every factor of the electric field $F^{i}_{x\tau}$ should appear as the combination $\sqrt{-g^{xx}g^{\tau\tau}}F^{i}_{x\tau}= e^i v_1^{-1}$. The magnetic field $F^m$ and the scalar field $\phi^i$ do not have any $v_1$ factors. There is no factor which comes from the covariant derivatives of the the aforementioned fields. The remaining factor comes from the overall multiplicative volume,
$\sqrt{-\mbox{det} g}$ which is proportional to $v_1$. So these make it possible to specify the function $L_\lambda$ in the following form,
\begin{equation} 
L_{\lambda}(\vec{e},\vec{p},\vec{u},\vec{v})=v_{1}\,g(\vec{e}v^{-1}_{1},\,\vec{p},\,\vec{u},\,\lambda v^{-1}_{1},\,v_{2}).
\end{equation}  

Now it is easy to see that
\begin{equation} 
v_{1}\frac{\partial L_{\lambda}}{\partial v_{1}}=L_{\lambda}-\vec{e}\cdot \frac{\partial L_{\lambda}}{\partial \vec{e}}-\lambda\frac{\partial L_{\lambda}}{\partial \lambda}.
\end{equation}  
Since the left hand side vanishes on shell, the entropy (\ref{entropy2}) of the extremal black hole solution will be
\begin{equation}\label{entropyfinal}
S_{BH}=2\pi \left(L- \vec{e}\cdot  \frac{\partial L}{\partial \vec{e} }\right)\equiv-2\pi H, 
\end{equation} 
where $H$ could be understood as the  `Routhian density' over two dimensional de Sitter space-time. 

We verify the above result by applying it to $4$-dimensional charged black holes of Nariai class.
Let us first consider the standard Einstein gravity coupled with the gauge fields in the presence of the positive cosmological constant in $d$-dimensions; 
\begin{equation}\label{actionRND}
\mathcal{S}=\int d^{d}x\sqrt{-g} \left[ \frac{1}{16\pi G_{d}}\left(R- 2\Lambda_{d} \right)- \frac{1}{4}\vert F^{(2)}_{e}\vert^{2}- \frac{1}{2\cdot (d-2)!}\vert F^{(d-2)}_{m}\vert^{2}  \right]. 
\end{equation}
The ans\"{a}tze (\ref{ansatz1}) and (\ref{ansatz2}) specify Lagrangian density (over $2$-dimensional de Sitter) into the form
\begin{eqnarray}\label{q14}
L&=&\int d\Omega_{d-2}\sqrt{-g}\,\mathcal{L}\\
&=&\mbox{Vol}(\mbox{S}^{d-2}) v_{1}v_{2}^{\frac{d-2}{2}}\left\{\frac{1}{16\pi G_{d}}\left( \frac{2}{v_{1}} + \frac{2}{v_{2}}-2\Lambda_{d}\right) + \frac{1}{2}\left( \frac{e^{2}}{v^{2}_{1}}- \frac{p^{2}}{\mbox{Vol}^{2}(\mbox{S}^{d-2}) v_{2}^{d-2}} \right) \right\}. \nonumber
\end{eqnarray}
The only difference of the above result from that of the extremal case is the plus sign of the term $2/v_{1}$.

The `Routhian density' $H$  will be expressed, in terms of the canonical conjugate momenta,
\begin{equation} 
\vec{q}=\frac{\partial L}{\partial \vec{e}}=\frac{\vec{e}}{v_{1}}v^{\frac{d-2}{2}}_{2}\mbox{Vol}(\mbox{S}^{d-2}),
\end{equation}  
as
 \begin{equation} 
H=\vec{q}\cdot \vec{e}-L=\frac{v_{1}\left(\vert\vec{q}\vert^{2}+\vert \vec{p}\vert^{2} \right) }{2 v^{\frac{d-2}{2}}_{2}\mbox{Vol}(\mbox{S}^{d-2})}-\frac{\mbox{Vol}(\mbox{S}^{d-2})}{8\pi G_{d}}\left( v^{\frac{d-2}{2}}_{2}+v_{1}v^{\frac{d-4}{2}}_{2}-\Lambda_{d}v_{1}v^{\frac{d-2}{2}}_{2}\right). 
\end{equation}  

In $4$-dimensions, the function $H$ becomes maximal at 
\begin{eqnarray} 
v_{1}&=&\frac{4\pi v^{2}_{2}}{4\pi v^{2}_{2}\Lambda_{4} -G_{4} \left(p^{2}+q^{2} \right) }, \nonumber\\
v_{2}&=&\frac{\pi+\sqrt{\pi^{2}-G_{4} \left(p^{2}+q^{2} \right)\pi\Lambda_{4} }}{2\pi\Lambda_{4}},
\end{eqnarray}
with the value
\begin{equation} 
H_{\mbox{max}}=-\frac{\pi+\sqrt{\pi^{2}-G_{4} \left(p^{2}+q^{2} \right)\pi\Lambda_{4}}}{4\pi G_{4}\Lambda_{4}}.
\end{equation}  

Though the value $H_{\mbox{max}}$ is negative, the entropy, as is obtained in (\ref{entropyfinal}), is positive;
\begin{equation} 
S_{BH}=-2\pi H_{\mbox{max}}=\frac{\pi+\sqrt{\pi^{2}-G_{4} \left(p^{2}+q^{2} \right)\pi\Lambda_{4}}}{2G_{4}\Lambda_{4}}.
\end{equation}    
This coincides with the horizon area divided by $4G_{4}$, if we set 
\begin{equation} 
(p^{2}+q^{2})=4\pi G_{4} Q^{2}.
\end{equation}    

A $4$-dimensional extremal Reissner-Nordstr\"{o}m-de Sitter black hole is characterized by the metric function
\begin{eqnarray}\label{rndsmetric}
f(r)&=&- \frac{1}{l^{2}r^{2}} \left(r-r_{0} \right)^{2} \left(r-r_{c} \right) \left( r+ \left(2r_{0}+r_{c} \right) \right)\nonumber\\
&=&1- \frac{2G_{4}M}{r}+ \frac{G_{4}^{2}Q^{2}}{r^{2}}-\frac{r^{2}}{l^{2}}.
\end{eqnarray}  
There are two possibilities of extremal cases, of which we are now interested in the case of $r_{0}>r_{c}$, that is when the double zero $r_{0}$ is larger than the simple zero $r_{c}$. The case corresponds to the charged black hole of Nariai class\footnote{The other case of $r_{c}>r_{0}$ corresponds to the charged extremal black hole in de Sitter background. See Ref. \cite{Cho:2007mn} for its detail.}. In its near-horizon, the geometry looks like a charged Nariai. The charged Nariai also appears as the geometry in the region between $r_{0}$ and $r_{c}$ in the extremal limit \cite{Bousso:1996au,Bousso:1996pn}.

Fig. \ref{figrn} illustrates the situation. Bekenstein-Hawking entropy, that is, the horizon area divided by $4G_{4}$ and is given by 
\begin{equation} 
S= \frac{\pi l^{2}}{6G_{4}}\left(1+\sqrt{1- \frac{12}{l^{2}}G_{4}^{2}Q^{2}} \right). 
\end{equation}  
Therefore we see that $S_{BH}=S$.


\FIGURE{
\centering
\includegraphics[width=9 cm]{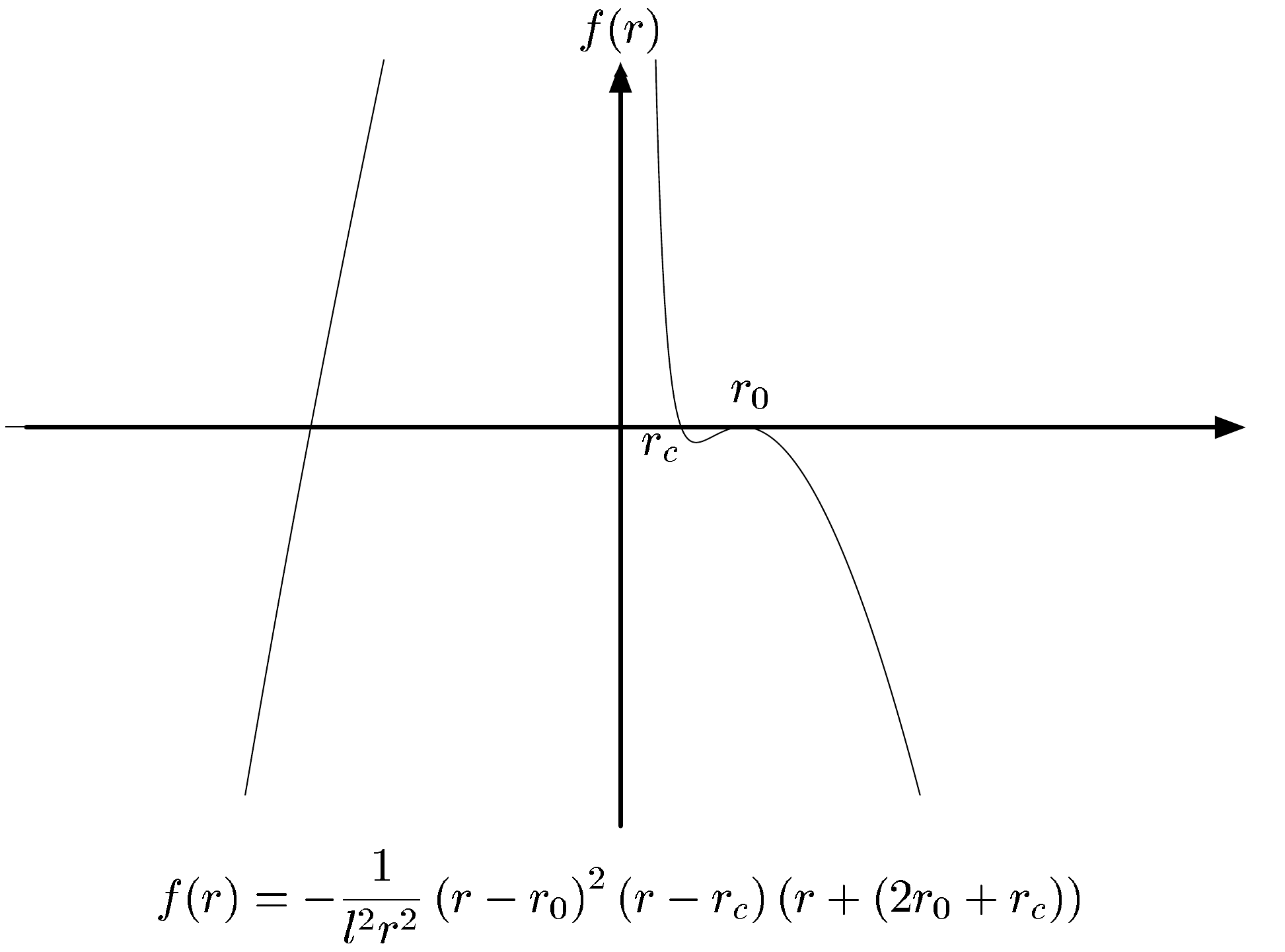}
\caption{\small The metric function $f(r)$ of $4$-dimensional extremal Reissner-Nordstr\"{o}m black hole has a double zero $r_{0}$ and a simple zero $r_{c}$. Only the case of $r_{0}>r_{c}$ corresponds to the black hole of Nariai class. The other case of $r_{c}>r_{0}$ is called a cold black hole and was discussed in Ref. \cite{Cho:2007mn}.}
\label{figrn}
}

\section{Gauss-Bonnet Corrections}\label{secv}

Let us consider Gauss-Bonnet corrections to the entropy function. With the coefficient $\alpha$ carrying the dimension of the length squared, Gauss-Bonnet term leads to the corrections in the $d$-dimensional action as 
\begin{equation}\label{higheraction}
\triangle \mathcal{S}= \frac{\alpha}{16\pi G_{d}}\int d^{d}x \sqrt{-g} \left(R^{MNPQ}R_{MNPQ}-4 R^{MN}R_{MN}+R^{2} \right). 
\end{equation} 

From the curvature components (\ref{curvature}), we obtain
\begin{eqnarray} 
R_{\alpha\beta}&=&\frac{1}{v_{1}}g_{\alpha\beta},\qquad R_{\mu\nu}= \frac{d-3}{v_{2}}g_{\mu\nu}, \nonumber\\
R&=&\frac{2}{v_{1}}+\frac{(d-2)(d-3)}{v_{2}}.
\end{eqnarray}
Inserting these results into Eq. (\ref{higheraction}), we get the following Lagrangian density over dS$_{2}$;
\begin{eqnarray}\label{lagaddition}
\triangle L&=&\frac{\alpha}{16\pi G_{d}}\mbox{Vol}(S^{d-2}) v_{1}v_{2}^{\frac{d-2}{2}} \left( \frac{1}{v^{2}_{2}}\left(d-4 \right) \left(d-5 \right) + \frac{4 }{v_{1}v_{2}}\right)  \left(d-2 \right) \left( d-3\right). 
\end{eqnarray} 
Compared to the case of adS$_{2}$, the second term comes with the opposite sign.

The term gives a non-trivial result for $d\ge 5$. In fact, in $d=4$, it leads to the following corrections to the entropy function
\begin{equation} 
\triangle S=-2\pi\triangle H=\frac{4\pi\alpha}{G_{4}}.
\end{equation}  
This contribution looks bizarre because it can be negative depending on the sign of $\alpha$. However, one should notice that the constant contribution is not concerned with any characteristic of the black hole. In fact, $\alpha$ is just the coefficient of Gauss-Bonnet term. Every black holes of Nariai class in the same theory will have this common constant contribution to the entropy. A reasonable interpretation is to view the entropy in the relative sense, which will trivialize the constant contribution. The same situation happens in the cold black holes, in which the double zero of the metric function is less than the simple zero \cite{Cho:2007mn}.
  
\section{Discussions}\label{secvi}

In this section, we conclude the paper by laying out two comments on the properties of our near-horizon Nariai geometry. First, we will explain how this near-horizon geometry is different from the one found in Refs. \cite{Ginsparg:1982rs,Bousso:1996au}. Second, we consider an issue concerning the  temperature of the black holes of Nariai class and suggest a way to resolve it. We argue that it should be non-zero despite the extremality of the black holes.
 
As was noted earlier in this paper, Nariai geometry appears also in the region between two horizons of a Schwarzschild-de Sitter black hole in the extremal limit \cite{Ginsparg:1982rs,Bousso:1996au}. However, it is different from the one we discussed in this paper. Though both geometries are locally the same, their global structures are different. They cover different portion of de Sitter space-time. The geometry discussed in the afore mentioned papers is the Euclidean version of Nariai geometry and corresponds to the following metric of the Lorentzian geometry:
\begin{equation}\label{ginsparg}
ds^{2}=\frac{l^{2}}{3}\left(-\sin^{2}{\chi}\,dt^{2}+d\chi^{2}+d\Omega^{2}_{2}  \right). 
\end{equation}    
It has {\it two different} horizons; the black horizon at $\chi=\pi$ and the cosmological horizon at $\chi=0$. Let us restrict our consideration to the two-dimensional de Sitter part. This static geometry does not cover the whole de Sitter space-time. One can relate the metric with more familiar form written in the conventional static coordinates using the relation
\begin{equation}\label{static}
r_{0}t=\zeta,\quad\,\sin{\chi}\,=\sqrt{1-\frac{r^{2}}{r^{2}_{0}}},  \qquad (\mbox{with }r^{2}_{0}= \frac{l^{2}}{3} )
\end{equation}  
as
\begin{equation} 
ds^{2}=-\left(1-\frac{r^{2}}{r^{2}_{0}} \right)d\zeta^{2}+ \left(1-\frac{r^{2}}{r^{2}_{0}} \right)^{-1}  dr^{2}+d\Omega^{2}_{2}. 
\end{equation}  
This latter form of the metric describes only one quarter of de Sitter space-time. However, the relation (\ref{static}) is a two-to-one mapping, the form of the metric (\ref{ginsparg}) therefore covers another twin partner. The situation is illustrated in Fig. \ref{twin}.


\FIGURE{
\centering
\includegraphics[width=9 cm]{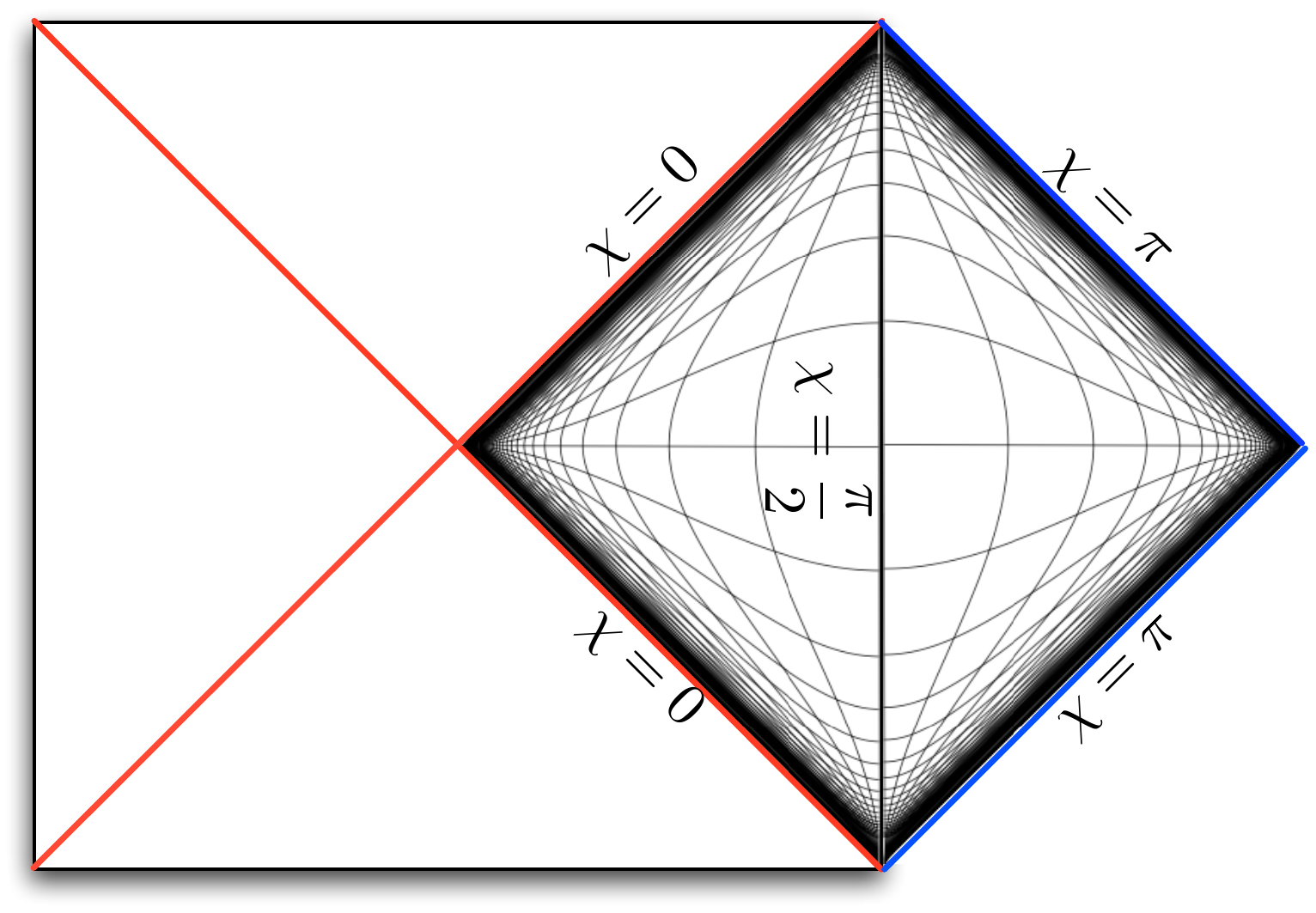}
\caption{\small Penrose diagram of the Nariai geometry between the black hole horizon (blue line at $\chi=\pi$) and the cosmological horizon (red line at $\chi=0$) of the Schwarzschild-de Sitter black hole in the extremal limit, where the geometry is symmetric under exchange of $\chi \leftrightarrow \pi-\chi$.}
\label{twin}
}

Meanwhile the Nariai geometry discussed in this paper involves only one horizon and the metric (\ref{planar}) written in the planar coordinates is time dependent and either coordinate patch $(\xi, y)$ or $(\lambda, y)$ covers half of the whole de Sitter space-time.

There is a temperature issue involved in the near horizon geometry of the black holes of Nariai class. The surface gravity read from the formula $\kappa=\vert f'(r_{0})/2\vert$ gives a null result for them because they are extremal in the sense that the event horizon coincides with the cosmological horizon. On the other hand, the observer living near the horizon of the black hole will definitely feel the temperature of two dimensional de Sitter space-time. It will be given by
\begin{equation}\label{temp1}
T=\frac{\kappa}{2\pi}=\frac{1}{2\pi l_{dS_{2}}},
\end{equation}  
where $l_{dS_{2}}$ is  the size of two-dimensional de Sitter space-time\footnote{The notation, $T$, of the temperature should not be confused with the one used in the earlier section for the trace of the energy-momentum tensor.}. In the extremal Schwarzschild-de Sitter black hole, $l_{dS_{2}}=r_{0}=l/\sqrt{3}$ where $l$ is the size of four-dimensional de Sitter space-time. In order to cure this discrepancy, one has to use the Killing vector of an observer following the geodesic line in computing the surface gravity \cite{Bousso:1996au}. Indeed the conventionally used Killing vector of the asymptotic observer does not make sense because there is no spatially asymptotic region in de Sitter background. However we still have a technical difficulty in applying the modified surface gravity (\ref{modified}) in computing the temperature
\begin{equation} 
\tilde{T}=\frac{\tilde{\kappa}}{2\pi}= \frac{\vert f'(r_{h})\vert}{4\pi\sqrt{-f(r_{g})}}
\end{equation}
of a black hole of Nariai class   
because at least one of the positions of the geodesic orbit is $r_{g}=r_{0}$ that makes $f(r_{g})=0$. Since the position of the horizon is also at $r_{h}=r_{0}$, we suggest the following limit value as the temperature of the black holes of Nariai class:
\begin{equation}\label{tempbar}
\bar{T}=\lim_{r \rightarrow r_{0}} \frac{\vert f'(r)\vert}{4\pi\sqrt{-f(r)}}=\frac{1}{2\pi}\sqrt{\frac{-f''(r_{0})}{2}}.
\end{equation}  
Indeed for the extremal Schwarzschild-de Sitter black hole, it gives $\bar{T}=1/2\pi l_{2}$ that is nothing but the temperature (\ref{temp1}). 

In the following, we will give more general argument that the above definition accords with the temperature read from the near horizon Nariai geometry. For general black holes of Nariai class, the metric function $f(r)$ can be expanded near the generate horizon in the near-horizon coordinate $\epsilon\tau=r-r_{0}$ as
\begin{equation} 
f(r)=\frac{f''(r_{0})}{2}\rho^{2}\epsilon^{2}+\mathcal{O}(\epsilon^{3}) 
\end{equation}  
and its value is mostly negative around the horizon. Therefore the near-horizon geometry takes the form of Nariai type:
\begin{equation} 
ds^{2}\simeq -\frac{1}{2}f''(r_{0})\tau^{2}x^{2}+ \frac{2}{f''(r_{0})}\frac{d\tau^{2}}{\tau^{2}}+r^{2}_{0}d\Omega^{2}_{d-2}
\end{equation}  
where was used the rescaled coordinate $x=\epsilon t$. Since the value of $f(r)$ is mostly negative around the degenerate horizon $r=r_{0}$, its second derive is negative at $r=r_{0}$, that is, $f''(r_{0})<0$. The size of two dimensional de Sitter space-time is 
\begin{equation} 
l^{2}_{dS_{2}}=-\frac{2}{f''(r_{0})}.
\end{equation}   
The temperature of two dimensional de Sitter space-time is given by
\begin{equation} 
T=\frac{1}{2\pi l_{dS_{2}}}=\frac{1}{2\pi}\sqrt{\frac{-f''(r_{0})}{2}}.
\end{equation}  
This temperature read from the near horizon Nariai geometry is coincident with the temperature $\bar{T}$ defined in (\ref{tempbar}).

Temperature issue also arises in the extremal black holes discussed in Ref. \cite{Cho:2007mn}. For example, the four-dimensional extremal Reissner-Nordstr\"{o}m-de Sitter black hole, though it has the same metric function $f(r)$ as in Eq. (\ref{rndsmetric}), has the degenerate horizon $r_{0}$ that is smaller than $r_{c}$. In the region between $r_{0}$ and $r_{c}$, there are two geodesic orbits to which the Killing vector $dx^{\mu}/dt$ is tangential; one is at $r_{0}$ and the other is at a point $r_{g}(\ne r_{0})$ inside the region. The temperature $\bar{T}$ measured by an observer at $r=r_{0}$ takes the same form as in (\ref{tempbar}), but without the minus sign inside the square root because $f''(r_{0})>0$ in the case. The result for the four-dimensional Reissner-Nordstr\"{o}m-de Sitter black hole is
\begin{eqnarray} 
\bar{T}&=& \frac{1}{2\pi l r_{0}}\sqrt{\pm \left(l^{2}-6r^{2}_{0} \right) }=\frac{1}{2G^{2}_{4}Q^{2}}\left(\sqrt{1- \frac{12}{l^{2}}G^{2}_{4}Q^{2}}\pm \left( 1- \frac{12}{l^{2}}G^{2}_{4}Q^{2}\right)   \right),
\end{eqnarray}
where the upper sign is for the extremal case and the lower one is for the case of Nariai class. The details about the expression for $r_{0}$ in terms of the charge $Q$ can be found in Ref. \cite{Cho:2007mn}. On the other hand, the temperature $\tilde{T}$ measured by an observer at $r_{g}$ is zero because $f'(r_{0})=0$. Regarding the extremal black holes in de Sitter background, one has to use this  temperature $\tilde{T}$ rather than the one $\bar{T}$. The vanishing temperature $\tilde{T}$ of the extremal black hole is consistent with its near horizon AdS$_{2}$ and with the geometry obtained in $l \rightarrow\infty$ limit. 

\acknowledgments{
We thank Yong-Wan Kim, Yun Soo Myung, and Young-Jai Park for stimulating discussions on Nariai geometry.
This work was supported by the SRC program of KOSEF through CQUeST with grant number R11-2005-021. It was also supported by the Korea Research Foundation Grant funded by the Korean Government(MOEHRD) (KRF-2007-314-C00056 ).
}

\end{document}